\documentclass{cupconf}

  \checkfont{eurm10}
  \iffontfound
    \IfFileExists{upmath.sty}
      {\typeout{^^JFound AMS Euler Roman fonts on the system,
                   using the 'upmath' package.^^J}%
       \usepackage{upmath}}
      {\typeout{^^JFound AMS Euler Roman fonts on the system, but you
                   dont seem to have the}%
       \typeout{'upmath' package installed. cupconf.cls can take advantage
                 of these fonts,^^Jif you use 'upmath' package.^^J}%
      }
  \else
  \fi

  \checkfont{msam10}
  \iffontfound
    \IfFileExists{amssymb.sty}
      {\typeout{^^JFound AMS Symbol fonts on the system, using the
                'amssymb' package.^^J}%
       \usepackage{amssymb}%
         \let\leq=\leqslant
         \let\geq=\geqslant
      }{}
  \fi

  \IfFileExists{amsbsy.sty}
    {\typeout{^^JFound the 'amsbsy' package on the system, using it.^^J}%
     \usepackage{amsbsy}}
    {}

\newsavebox{\astrutbox}
\sbox{\astrutbox}{\rule[-5pt]{0pt}{20pt}}

\input psfig.sty
\title[Wolf-Rayets at High Metallicity]{Wolf-Rayet Populations at
High Metallicity}

\author[P. A. Crowther]{P\ls A\ls U\ls L\ns A.\ns C\ls R\ls O\ls W\ls T\ls H\ls E\ls R$^1$}

\affiliation{$^1$Department of Physics \& Astronomy, University of Sheffield,
Hicks Building, Hounsfield Road, Sheffield, S3 7RH, UK}

\pubyear{2006}
\volume{XXX}
\pagerange{XXX}
\date{?? and in revised form ??}
\begin{document}

\maketitle

\begin{abstract} Observed properties of Wolf-Rayet stars at high
metallicity are reviewed. Wolf-Rayet stars are more common at higher
metallicity, as a result of stronger mass-loss during earlier evolutionary
phases with late WC subtypes signatures of solar metallicity or higher.
Similar numbers of early 
(WC4--7) and late (WC8--9) stars are observed in the Solar neighbourhood, 
whilst late subtypes dominate at higher metallicities, such
as Westerlund~1 in the inner Milky Way and in M83. The
observed trend to later WC subtype within metal-rich environments is 
intimately linked to a metallicity
dependence of WR stars, in the sense that strong winds preferentially
favour late subtypes. This has relevance to (a) the upper mass limit in
metal-rich galaxies such as NGC~3049, due to softer ionizing fluxes from
WR stars at high metallicity; (b) evolutionary models including a WR
metallicity dependence provide a better match to the observed N(WC)/N(WN)
ratio. The latter conclusion partially rests upon the assumption of
constant line luminosities for WR stars, yet observations and theoretical
atmospheric models reveal higher line fluxes at high metallicity.
\end{abstract}

\firstsection % if your document starts with a section,
\section{Introduction}

Wolf-Rayet (WR) stars represent the final phase in the evolution
of very massive stars prior to core-collapse, in which the H-rich
envelope has been stripped away via either stellar winds or close
binary evolution, revealing products of H-burning (WN sequence) or
He-burning (WC sequence) at their surfaces, i.e. He, N or C, O (Crowther
2007). 

\begin{figure}
\centerline{\psfig{file=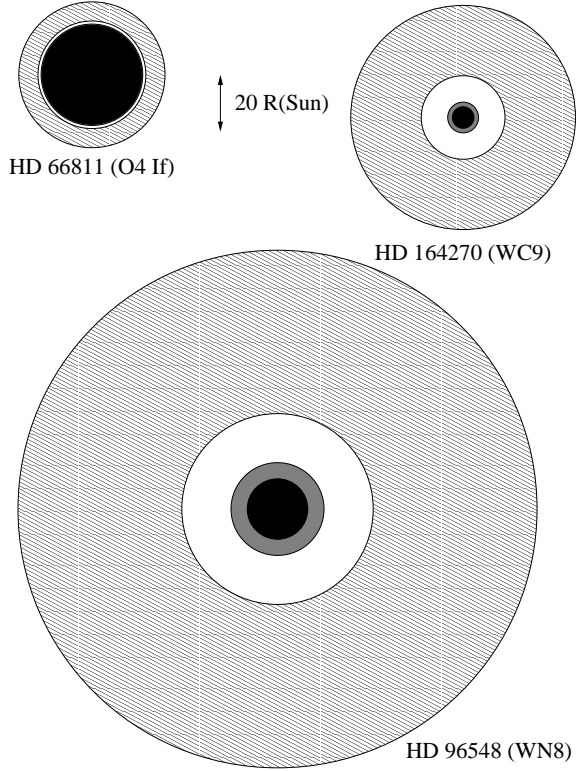,width=3.0in}}
\caption{Comparisons between stellar radii at Rosseland
optical depths of 20 (= $R_{\ast}$, black) and 2/3 (= $R_{2/3}$, grey)
for HD~66811 (O4\,If), HD~96548 (WN8) and HD~164270 (WC9), shown to
scale, together with the wind region corresponding to the primary
optical wind line forming region, $10^{11} \leq n_{e} \leq 10^{12}$
cm$^{-3}$ (hatched) in each case, illustrating the highly extended
winds of WR stars with respect to O stars (Crowther 2007).}\label{WRross}
\end{figure}

WR stellar winds are significantly denser than O stars, as illustrated in
Fig.~\ref{WRross}, so their visual spectra are dominated by broad emission
lines, notably He\,{\sc ii} $\lambda$4686 (WN stars) and C\,{\sc iii} 
$\lambda$4647-51,
C\,{\sc iii} $\lambda$5696, C\,{\sc iv} $\lambda$5801-12 (WC stars). The 
spectroscopic
signature of WR stars may be seen individually in Local Group galaxies
(e.g. Massey \& Johnson 1998), within knots in local star forming galaxies
(e.g. Hadfield \& Crowther 2006) and in the average rest frame UV spectrum
of Lyman Break Galaxies (Shapley et al. 2003).

In the case of a single massive star, the strength of stellar winds during
the main sequence and blue supergiant phase scales with the metallicity
(Vink et al. 2001). Consequently, one expects a higher threshold for the
formation of WR stars at lower metallicity, and indeed the SMC shows a
decreased number of WR to O stars than in the Solar Neighbourhood.  
Alternatively, the H-rich envelope may be removed during the Roche lobe
overflow phase of close binary evolution, a process which is not expected
to depend upon metallicity.

WR stars represent the prime candidates for Type Ib/c core-collapse
supernovae and long, soft Gamma Ray Bursts (GRBs).  This is due to their
immediate progenitors being associated with young massive stellar
populations, compact in nature and deficient in either hydrogen (Type Ib)
or both hydrogen and helium (Type Ic).  For the case of GRBs, a number of
which have been associated with Type Ic hypernovae (Galama et al. 1998;
Hjorth et al. 2003), a rapidly rotating core is a requirement for the
collapsar scenario in which the newly formed black hole accretes via an
accretion disk (MacFadyen \& Woosley 1999).  Indeed, WR populations have
been observed within local GRB host galaxies (Hammer et al. 2006).

In this review article, the observed properties WR stars at high 
metallicity are presented and discussed.

\begin{figure}
\centerline{\psfig{file=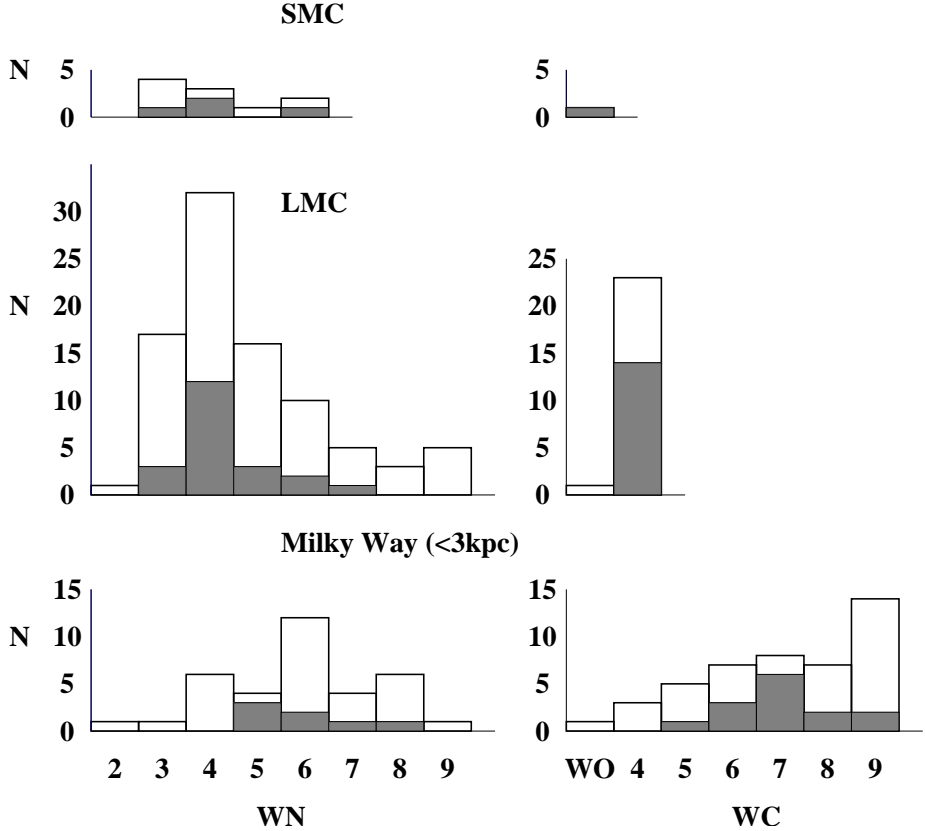,width=4.8in}}
\caption{Subtype distribution of Milky Way ($<$3kpc),
LMC and SMC WR stars, in which known binaries are
shaded (Crowther 2007).}\label{wrpop}
\end{figure}

\section{WR subtype distribution in Milky Way and Magellanic Clouds}

Historically, the wind properties of WR stars have been assumed to be
metallicity independent (Langer 1989), yet there is a well known
observational trend to later, lower ionization, WN and WC subtypes at
high metallicity as illustrated in Fig.~\ref{wrpop}.
%
% Mass-loss rates for WN stars in the Milky Way and LMC show a very large
% scatter. The presence of hydrogen in some WN stars further complicates 
% the
% picture since WR winds are denser if Hydrogen is absent (Nugis \& Lamers 
% 2000). 
% 

Within the Milky Way, it is well known that late-type WC stars are
restricted to within the Solar circle (Conti \& Vacca 1990). Indeed, 
Hopewell et al. (2005) discovered five new WR stars in the inner Milky
Way using the AAO/UKST H$\alpha$ survey -- all were found to be  WC9 
stars. 

In addition, Westerlund~1 (Clark et al. 2005) -- located at 
the edge of the Galactic bar, for which a metallicity of 
$\sim$60\% higher than Orion is expected -- possesses 8 WC stars 
with a bias towards late (WC8--9) subtypes according to recent near-IR 
spectroscopy (Crowther et al. 2006). The majority of these possess hot
dust, indicative of massive binaries, in common with the Quintuplet 
members of the Galactic Centre Quintuplet cluster (Figer, priv. comm.).
Early-type WN stars are also absent in Westerlund~1, with equal numbers of 
mid (WN5--6) and late (WN7--10) subtypes, for which the majority also 
appear  to be  massive binaries, as a result of hard X-ray fluxes.

\begin{figure}
\centerline{\psfig{file=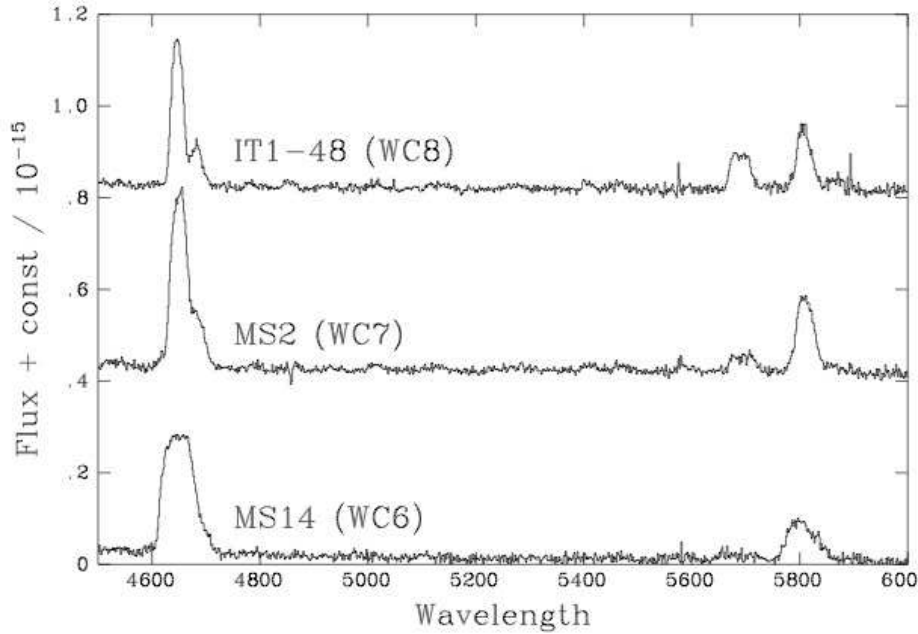,width=4.8in,angle=-90}}
\caption{Optical WHT/ISIS spectroscopy of representative WC stars
in M31.}\label{m31}
\end{figure}

\section{WR populations in M31, M83 and beyond}

Within the Local Group, M31 (Andromeda) is the only other 
candidate metal-rich  galaxy, although studies are hindered by its 
orientation on the sky.
Wolf-Rayet populations in M31 were led by Moffat \& Shara (1983, 1987)
and Massey et al. (1986). As in the Milky Way, WC7--8 stars 
were located at smaller galacto-centric distances (7$\pm$3 kpc) than
WC5--6 stars (11$\pm$3 kpc). Representative examples of M31 WC stars
obtained with WHT/ISIS are presented in Fig.~\ref{m31}, and suggest
a weak metallicity-gradient for M31, albeit rather less metal-rich than
the Milky Way, according to its observed WC population (see also
Trundle et al. 2002).

% WHT/ISIS figure? IT1-48 (1 WC8), MS2 (5 WC7), MS14 (3 WC6)

Further afield, the Wolf-Rayet population of M83 (= NGC~5236)  has been
studied by Hadfield et al. (2005). M83 is 
well suited to optical imaging surveys for WR stars at high 
metallicity (though see Bresolin, these proceedings), since it is face-on,
nearby (4.5~Mpc), and possesses a high star formation rate. 

\begin{figure}
\centerline{\psfig{file=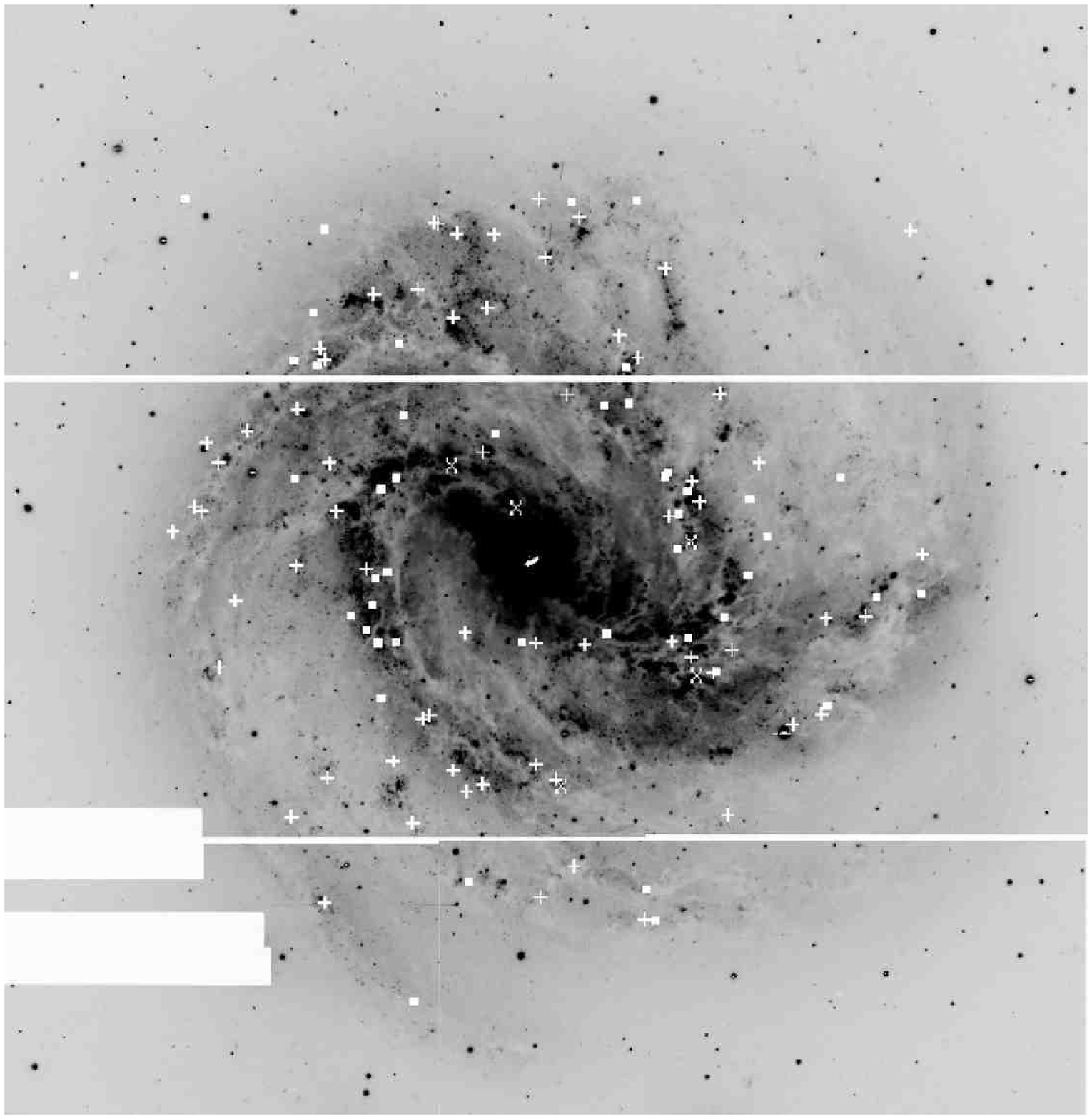,width=4.8in,angle=0}}
\caption{Composite 12$' \times 12'$ VLT FORS2 image 
($\lambda$4685) of M83 (= NGC~5236) indicating the location
of regions containing  WN (squares), WC (plus symbols) and WN+WC
(crosses) stars. North is up and east is to the left. Refions
to the south east are masked to avoid saturation by bright 
foreground stars (Hadfield et al. 2005)}\label{m83}
\end{figure}

\begin{figure} 
\centerline{\psfig{file=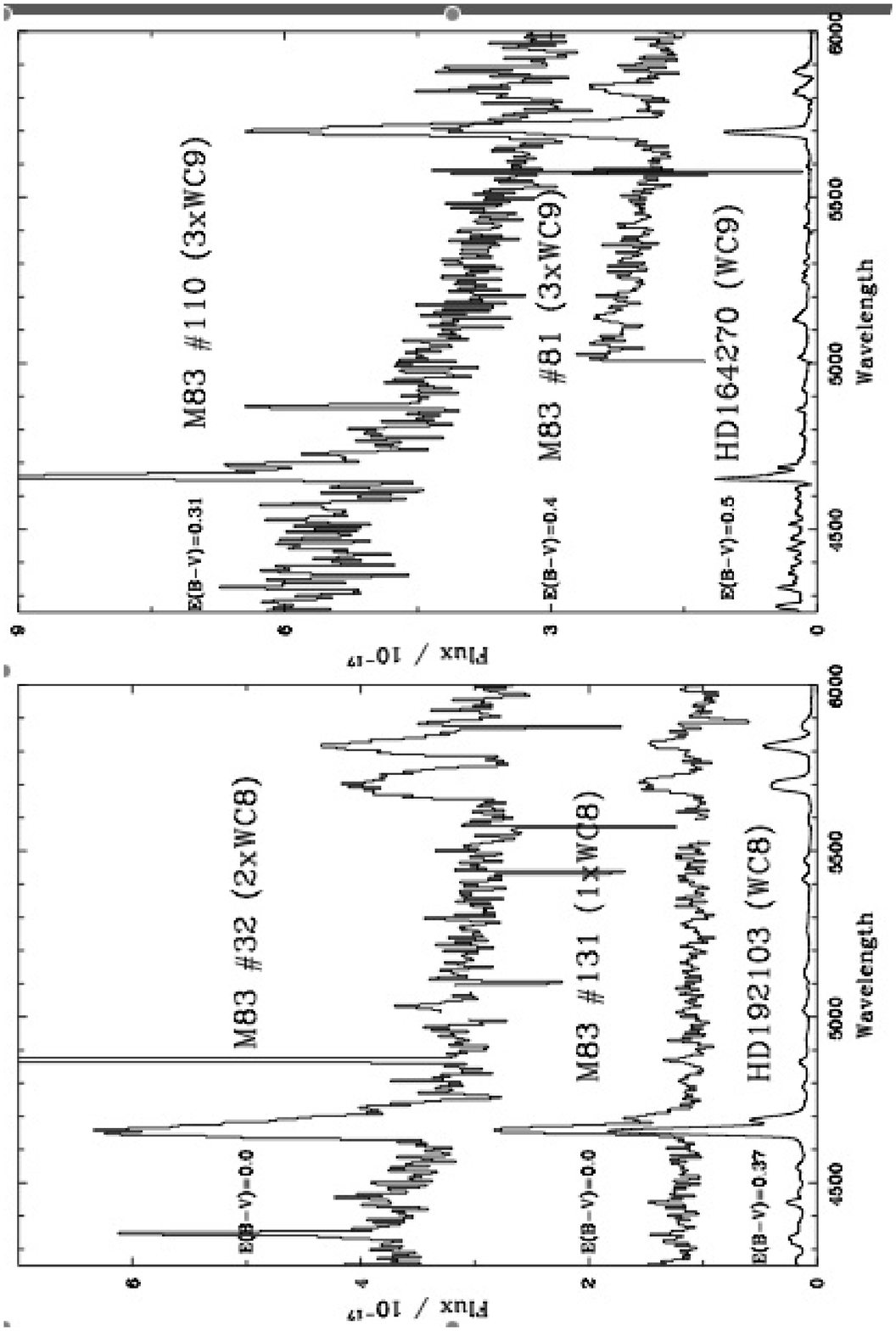,width=4.8in,angle=-90}}
\caption{Optical VLT/FORS2 spectroscopy of typical regions of M83 hosting 
late WC8--9 stars, 
together with individual Milky Way WC stars (WR135, WR103)
scaled to a distance of 4.5Mpc (Hadfield et al. 2005).}\label{m83_wcl} 
\end{figure}

VLT/FORS2  revealed a total of 280 (non-nuclear) 
regions for which the presence of WR stars was inferred from an excess
of $\lambda$4685 (He\,{\sc ii} 4686) narrow-band imaging versus 
$\lambda$4781 
(continuum) -- see Fig.~\ref{m83}. Spectroscopic follow-up of 198 regions 
confirmed a total  of 132 sources, hosting in excess of 1000 WR stars 
according to  the standard calibration of Schaerer \& Vacca (1998). 
N(WC)/N(WN)$\sim$1.2 in M83, confirming the trend towards a higher ratio 
at high metallicity as indicated in Fig.~\ref{wc_wn}. Notably, the WC 
population of M83 is  totally dominated by late subtypes, i.e. N(WC8--9) / 
N(WC4--7) $\sim$9, in  contrast with $\sim$1 for the Solar neighbourhood,
for which representative examples are presented in Fig.~\ref{m83_wcl}.

WR populations are unresolved in more remote metal-rich galaxies, but also
confirm the presence of late-type WC stars, as originally discovered by
Phillips \& Conti (1992) for NGC~1365 within Fornax, and confirmed by
Pindao et al. (2002) for NGC~4254 (= M88) within Virgo.

\section{Why late WC stars at high metallicity?}

The ubiquitous detection of late WC stars within metal rich environments
(and absence at low metallicity) requires explanation. 
The observed trend for WC subtypes in the LMC versus the Milky Way was 
initially believed  to originate from a difference in carbon 
abundances (Smith \& Maeder 1991), yet  quantitative analysis
reveals similar carbon abundances (Koesterke \& Hamann 1995; Crowther et
al. 2002). Alternatively, late WC stars might evolve preferentially
from relatively low mass OB stars that enter the WR phase owing to
stronger stellar winds at earlier evolutionary phases. This scenario
is not supported by cluster studies (e.g. Massey et al. 2000; Crowther
et al. 2006).

The most compelling evidence suggests that late WC stars are favoured
in the case of high wind densities (Crowther et al. 2002). Consequently,
the presence of late-WC stars within metal-rich galaxies favours 
metallicity dependent winds for Wolf-Rayet stars. 
The impact of a metallicity dependence for WR winds upon spectral
types is as follows. At high metallicity, recombination from high
to low ions (early to late subtypes) is very effective in very
dense winds, whilst the opposite is true for low metallicity, low
density winds.  Stellar temperatures  further complicates this picture, 
such that the spectral type of a WR star  results from a subtle 
combination of  ionization and wind density, in contrast with normal stars.

Theoretically, Nugis \& Lamers (2002) argued that the iron opacity peak
was the origin of the wind driving in WR stars, which Gr\"{a}fener \&
Hamann (2005) supported via an hydrodynamic model for an early-type WC
star in which lines of Fe IX-XVII deep in the atmosphere provided the
necessary radiative driving. Vink \& de Koter (2005) applied a Monte Carlo
approach to investigate the metallicity dependence for cool WN and WC
stars revealing $\dot{M} \propto Z^{\alpha}$ where $\alpha$=0.86 for WN
stars and $\alpha$=0.66 for WC stars for 0.1 $\leq Z \leq 1 Z_{\odot}$.
The weaker WC dependence originates from an increasing Fe content and
constant C and O content at high metallicity.  Empirical results for the
Solar neighbourhood, LMC and SMC are broadly consistent with theoretical 
predictions, although detailed studies of  individual WR  stars within 
galaxies broader range in metallicity would provide stronger constraints.

\begin{figure}
\centerline{\psfig{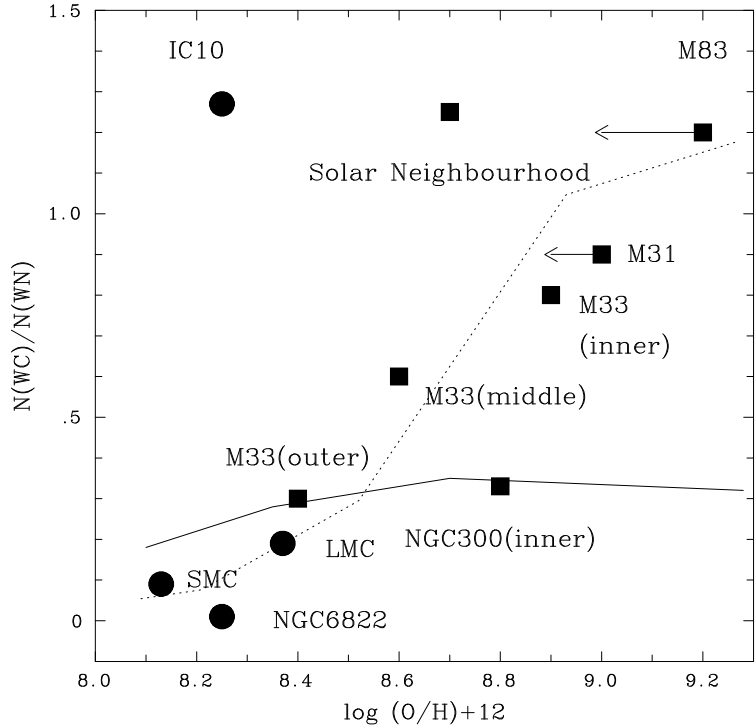}}
\caption{Ratio of subtype distribution of WC to WN stars
for nearby galaxies, updated from Massey \& Johnson (1998)
to include M83 (Hadfield et al. 2005). Evolutionary 
predictions from Meynet \& Maeder (2005, solid line) and Eldridge \& 
Vink (2006, dotted line) are included.}\label{wc_wn}
\end{figure}

A metallicity dependence of WR winds impacts upon evolutionary model
calculations as follows. Recent evolutionary models of Meynet \& Maeder 
(2005) allow for rotational  mixing, but not a metallicity dependence of 
WR winds. Improved agreement with respect to earlier models is achieved,
but the ratio of WC versus WN stars for continuous star formation does
does not reproduce that observed at high metallicity, as illustrated in
Fig.~\ref{wc_wn}. In contrast, recent (non-rotating) evolutionary models 
by Eldridge \&  Vink (2006) in which the Vink \& de Koter (2005) WR 
metallicity dependence has been implemented provide a much better match
to observations.

With regard to the inferred WR populations at high metallicity a note of
caution is necessary. At high metallicity, WR optical recombination lines
will (i)  increase in equivalent width, since their strength scales with
the square of the density, and (ii) increase in line flux, since the lower
wind strength will reduce the line blanketing, resulting in an increased
extreme UV continuum strength at the expense of the UV and optical
(Crowther \& Hadfield 2006). Indeed, the equivalent widths of optical
emission lines of WN stars in the Milky Way and LMC WN stars are well
known to be higher than SMC counterparts (Conti et al.  1989).

To date, the standard approach for the determination 
of unresolved WR populations in external galaxies has been to assume 
metallicity  independent WR line fluxes --  obtained for Milky Way and LMC 
stars (Schaerer  \& Vacca 1998) -- regardless of whether the host galaxy 
is metal-rich (Mrk 309, Schaerer et al. 2000) or metal-poor (I~Zw\,18, 
Izotov et al. 1997). Ideally, one would wish to use WR template stars 
appropriate to the
metallicity of the galaxy under consideration.  Unfortunately, this is
only feasible for the LMC, SMC and Solar neighbourhood, since it is
challenging to isolate individual WR stars from ground based observations
in more distant galaxies, which span a larger spread in metallicity.

Enhanced WR line fluxes are also predicted for WR atmospheric models at
high metallicity if one follows the metallicity dependence from Vink \& de
Koter (2005), such that WR populations inferred from Schaerer \& Vacca
(1998) at high metallicity may overestimate actual populations. 

\begin{figure}
\centerline{\psfig{file=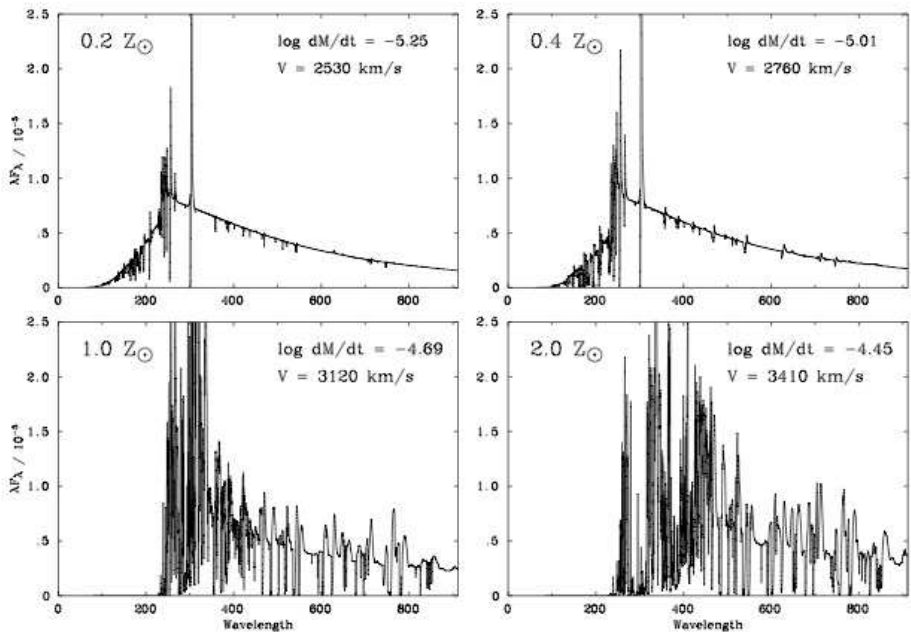,width=4.8in,angle=-90}}
\caption{Predicted Lyman continuum ionizing fluxes for
model reference WN\#11
($T_{\ast}$=100kK, $L=10^{5.48} L_{\odot}$) from Smith et al. 
(2002), illustrating
harder ionizing fluxes at lower metallicity, notably below 
$\lambda$=228\AA\ due to weaker stellar winds.}\label{ion}
\end{figure}

\section{Impact on ionizing fluxes}

Indirect H\,{\sc ii} region studies have suggested a low upper mass 
limit, $M_{\rm upper}$, for H\,{\sc ii} regions at high metallicity (e.g.
Thornley et al. 2000), yet Schaerer et al. (2000) claim $M_{\rm upper} 
> 40 M_{\odot}$ due to large WR populations in high metallicity galaxies
such as Mrk~309. However, Schmutz et al. (1992) demonstrated that the 
ionizing fluxes from WR stars
soften as wind density increases. Consequently, a metallicity dependence
for WR wind strengths implies that WR ionizing flux distributions soften
at increased metallicity, as demonstrated by Smith et al. (2002), and
naturally resolves this apparent discrepancy. For example, high 
temperature WN models predict a strong ionizing flux shortward of 
the He\,{\sc ii} Lyman edge at 228\AA\ at low metallicities, but 
negligible hard ionizing fluxes at Solar metallicities or above, as
illustrated in Fig.~\ref{ion}.

To illustrate this, we use the example of the metal-rich WR galaxy 
NGC~3049 (Schaerer et al. 1999). Gonzalez Delgado et al. (2002) identified 
a compact nuclear starburst cluster as the principal origin of WR emission
within NGC~3049. A nebular analysis based upon the hard
(metallicity-independent) WR ionizing fluxes of Schmutz et al. (1992) 
confirmed  previous results, i.e. $M_{\rm upper} < 40 M_{\odot}$. In 
contrast, use of the revised (metallicity dependent), softer WR ionizing
fluxes from Smith et al. (2002) led to a normal, $M_{\rm upper} \geq 100 
M_{\odot}$, in agreement with UV spectral synthesis studies.

\section{Summary}

Wolf-Rayet stars are more common at higher metallicity, as a result of
stronger mass-loss during earlier evolutionary phases. Late WC subtypes
appear to be signatures of solar metallicity or higher, as witnessed 
within the Westerlund~1 and Quintuplet clusters in the inner Milky Way,
M31, M83, NGC~3049, Mrk~309. The observed trend to later WC subtype is 
intimately linked to a metallicity dependence of WR stars, in the sense
that strong winds preferentially favour late subtypes (Crowther et al. 
2002). This has relevance to (a) the upper mass limit in metal-rich 
galaxies, due to softer ionizing 
fluxes from WR stars at high metallicity (Gonzalez Delgado et al. 2002);
(b) evolutionary models including a WR metallicity dependence provide
a better match to the observed N(WC)/N(WN) ratio (Eldridge \& Vink 2006).
The latter item relies in part upon the assumption of constant line
luminosities for WR stars (e.g. Schaerer \& Vacca 1998), yet observations 
and theoretical atmospheric models reveal higher line fluxes at high 
metallicity (Crowther \& Hadfield 2006).

\begin{acknowledgements}
Many thanks to Lucy Hadfield, with whom the majority of the results presented
here on M83 and Westerlund~1 were obtained, plus Chris Evans who 
carried out the M31 WHT/ISIS observing run. PAC acknowledges financial support from the Royal Society.
\end{acknowledgements}

\end{document}